\documentclass[reprint,superscriptaddress, amsmath,amssymb,aps,prl]{revtex4-2}

\usepackage{amssymb}
\usepackage{amsmath}
\usepackage{graphicx}
\usepackage{bm}
\usepackage{color}
\usepackage{hyperref}
\usepackage[version=4]{mhchem}
\hypersetup{
    colorlinks=true,
    linkcolor=blue,
    filecolor=magenta,      
    urlcolor=cyan,
    pdftitle={manuscript},
    pdfpagemode=FullScreen,
    }

\newcommand{\ICl}{$\beta'$-ET$_2${\rm ICl}$_2$}

\begin{document}
\title{Pressure-Induced Topological Changes in Fermi Surface \\
of Two-Dimensional Molecular Conductor} 

\author{T.~Kobayashi}
 \email{tkobayashi@mail.saitama-u.ac.jp}
 \affiliation{Graduate School of Science and Engineering, Saitama University, Saitama 338-8570, Japan}
 \affiliation{Research and Development Bureau, Saitama University, Saitama 338-8570, Japan}
  \author{K.~Yoshimi}
 \email{k-yoshimi@issp.u-tokyo.ac.jp}
  \author{H.~Ma}
 \affiliation{Institute for Solid State Physics, University of Tokyo, Kashiwa, Chiba 277-8581, Japan}
 \affiliation{Beijing National Laboratory for Condensed Matter Physics and Institute of Physics, Chinese Academy of Sciences, Beijing 100190, China}
 \author{S.~Sekine}
  \affiliation{Graduate School of Science and Engineering, Saitama University, Saitama 338-8570, Japan}
\author{H.~Taniguchi}
 \affiliation{Graduate School of Science and Engineering, Saitama University, Saitama 338-8570, Japan}
\author{N.~Matsunaga}
\author{A.~Kawamoto}
 \affiliation{Department of Condensed Matter Physics, Graduate School of Science, Hokkaido University, Sapporo 060-0810, Japan}
  \author{Y.~Uwatoko}
 \affiliation{Institute for Solid State Physics, University of Tokyo, Kashiwa, Chiba 277-8581, Japan}
\date{\today}

\begin{abstract}
We demonstrated X-ray structural analysis of the pressure-induced superconductor, $\beta'$-ET$_2${\rm ICl}$_2$ under extremely high pressure conditions, where ET denotes bis(ethylenedithio)tetrathiafulvalene. This material has been known as the highest transition temperature ($T_c$) superconductor among organic superconductors ($T_c$=14.2~K at 8.2~GPa).
On the basis of the experimental results, \textit{ab-initio} models were derived using the constrained random phase approximation.
We revealed that the Lifshitz transition exists behind the Mott insulator--imetal transition and found that the value of the on-site Coulomb interaction was halved to around $10$~GPa compared to that at ambient pressure. 
This study clarifies the enigmatic origins of high $T_{\rm c}$, and concurrently, provides a new understanding of the impacts of structural alterations in organic materials under high pressure on their electronic properties and the superconductivity process.
\end{abstract}

\maketitle

{\it Introduction---.}
Electron correlations give rise to diverse functional properties such as ferromagnetism, high-temperature superconductivity, and multiferroics \cite{Imada1998,Magda2014,Keimer2015,Tokura2017}. 
Theoretical prediction of novel material properties and subsequent synthesis of materials accordingly represent one of the research approaches that should be targeted in materials science.
Achieving this goal requires a microscopic understanding of the mechanisms and the subsequent ability to control them. 

Among strongly correlated electron systems, organic molecular solids stand out, exhibiting various quantum phases like Mott insulator \cite{Lefebvre2000,Limelette2003,Kagawa2005,Kagawa2009}, charge ordering \cite{Seo2000,Miyagawa2000,Kagawa2013}, antiferromagnetic ordering \cite{Welp1992,Miyagawa1995,Ishikawa2018,Oinuma2020}, and superconducting (SC) states \cite{Jerome1991,Williams1991,Coronado2000,Uji2001,Lortz2007}.
Characterized by crystal structures held together by van der Waals forces, organic molecular solids exhibit greater flexibility than inorganic counterparts, readily undergoing volume changes by applying pressure. 
These volume changes impact the transfer integrals, which in turn influence many-body quantum effects. 
Leveraging this feature, new quantum phases have been discovered by controlling physical properties through the application of pressure \cite{Murata1985a,Katayama2006,Itoi2008,Kato2017,Hirata2017,Samanta2021}.

\begin{figure}[t] 
\begin{center} 
\includegraphics[width=0.95\columnwidth]{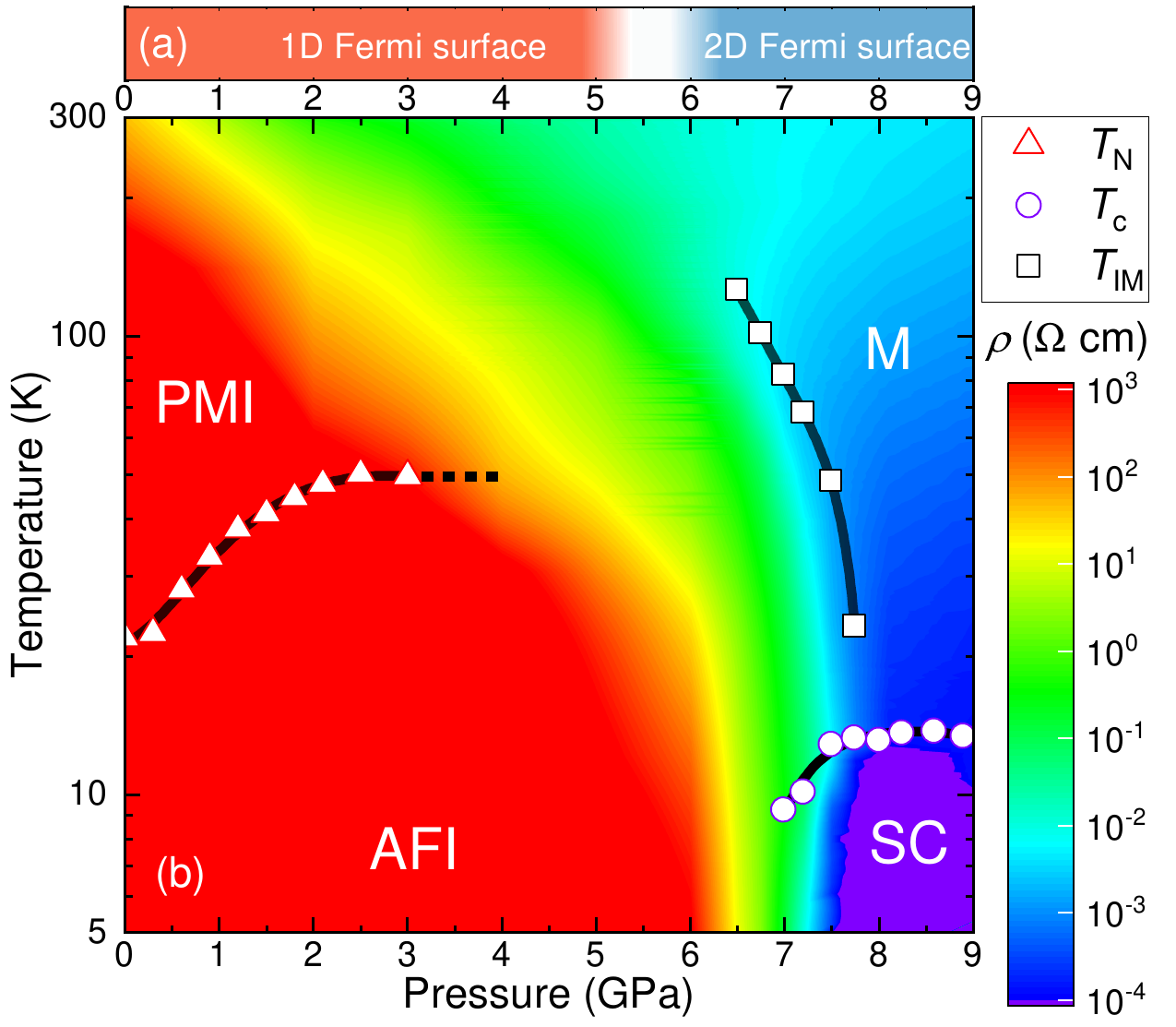}
\caption{
(a) Pressure evolution of shape of Fermi surface of \ICl\ revealed by present study. 
(b) Pressure--temperature contour plot of resistivity $\rho$ of \ICl  based on reported results \cite{Taniguchi2003a}, where $T_{\rm N}$,  $T_{\rm IM}$, and $T_{\rm c}$ are the N\'{e}el temperature \cite{Eto2010}, insulator--metal crossover temperature, and onset temperature of the SC transition, respectively. 
The solid lines are guides for the eye, and $T_{\rm N}$ above $3$~GPa has not been determined (dotted line).
} 
\label{fig-PT} 
\end{center}
\end{figure}

\ICl, [ET: bis(ethylenedithio)tetrathiafulvalene], is known to have the highest SC transition temperature $T_{\rm c}$ among organic conductors \cite{Taniguchi2003a}. 
As shown in the pressure--temperature phase diagram in Fig.~\ref{fig-PT}(b), the application of ultrahigh pressure induces a crossover from a paramagnetic insulating (PMI) phase to a metallic (M) phase, and at low temperatures an SC phase is adjacent to an antiferromagnetic insulating (AFI) phase \cite{Yoneyama1997,Eto2010}.
To clarify the origin of these phenomena, determining the high-pressure crystal structure is of primary importance. However, due to the large number of atoms in organic materials and the fragility of their crystals, it is quite difficult to determine atomic positions of organic materials by single-crystal X-ray diffraction measurements under pressure as high as $10$~GPa \cite{Tidey2014}. 

Instead, first-principles calculations were used to estimate the atomic positions under pressure by optimizing the internal coordinates and lattice parameters from ambient-pressure crystal structure and to calculate the electronic structure \cite{Kobayashi1986,Miyazaki2003}.
On the basis of the obtained electronic states, a theoretical analysis using the fluctuation exchange approximation suggested a possible link between a large density of states near the Fermi energy and a high $T_{\rm c}$ \cite{Kino2004,Nakano2006}.
However, the pressure range over which this mechanism is valid is much larger than in experiments \cite{Taniguchi2003a}, and the values of the electron--electron Coulomb repulsion $U$, including its pressure dependence, are also unknown, so the origin of superconductivity remains a mystery.

In this Letter, we successfully determined the atomic position of \ICl\ up to $\sim 10$~GPa, which includes the SC phase, by single-crystal X-ray structural analysis.
On the basis of the obtained crystal structures, we numerically derived the low-energy effective \textit{ab-initio} models. 
We reveal that the Lifshitz transition \cite{Lifshitz1960} exists behind the Mott insulator--metal transition and quantitatively show the pressure dependence of the electron correlations. 
The results show that we have successfully solved the puzzle of the gap between experiment and theory and have shed light on the mechanism of the onset of the SC state.

{\it Experimental---.}
Single crystals of \ICl\ were prepared by an electrochemical oxidation method \cite{Emge1986,Taniguchi2006}. 
X-ray diffraction data were collected at $293$~K using a Rigaku XtaLAB diffractometer with a MicroMax-007 HF microfocus rotating anode X-ray source (Mo-K$_{\alpha}$ radiation, $\lambda$ = $0.71073$~\AA) and a HyPix-6000 detector. 
The initial structure was solved by direct methods (SHELXT) \cite{Sheldrick2015a} and a structural refinement was performed by full-matrix least-squares methods based on $F^2$ (SHELXL) \cite{Sheldrick2015} with the Olex2 program \cite{Dolomanov2009}. 
High-pressure diffraction experiments were carried out using a diamond anvil cell (DAC) with a 0.3-mm culet size and a rhenium metal gasket.
A single crystal with a size of $0.15 \times 0.10 \times 0.03$~mm$^3$ was mounted in a DAC together with a piece of ruby as a pressure manometer. 
A series of high-pressure experiments were performed on this single crystal.
A methanol--ethanol mixture (4:1) with a solidification pressure of $10.5$~GPa \cite{Klotz2009} was used as the pressure-transmitting medium.
Ultrahigh pressure application under hydrostatic conditions was achieved except for $\sim 11$~GPa, the highest pressure reached in this study.

\begin{figure}[t] 
\begin{center} 
\includegraphics[width=1\columnwidth]{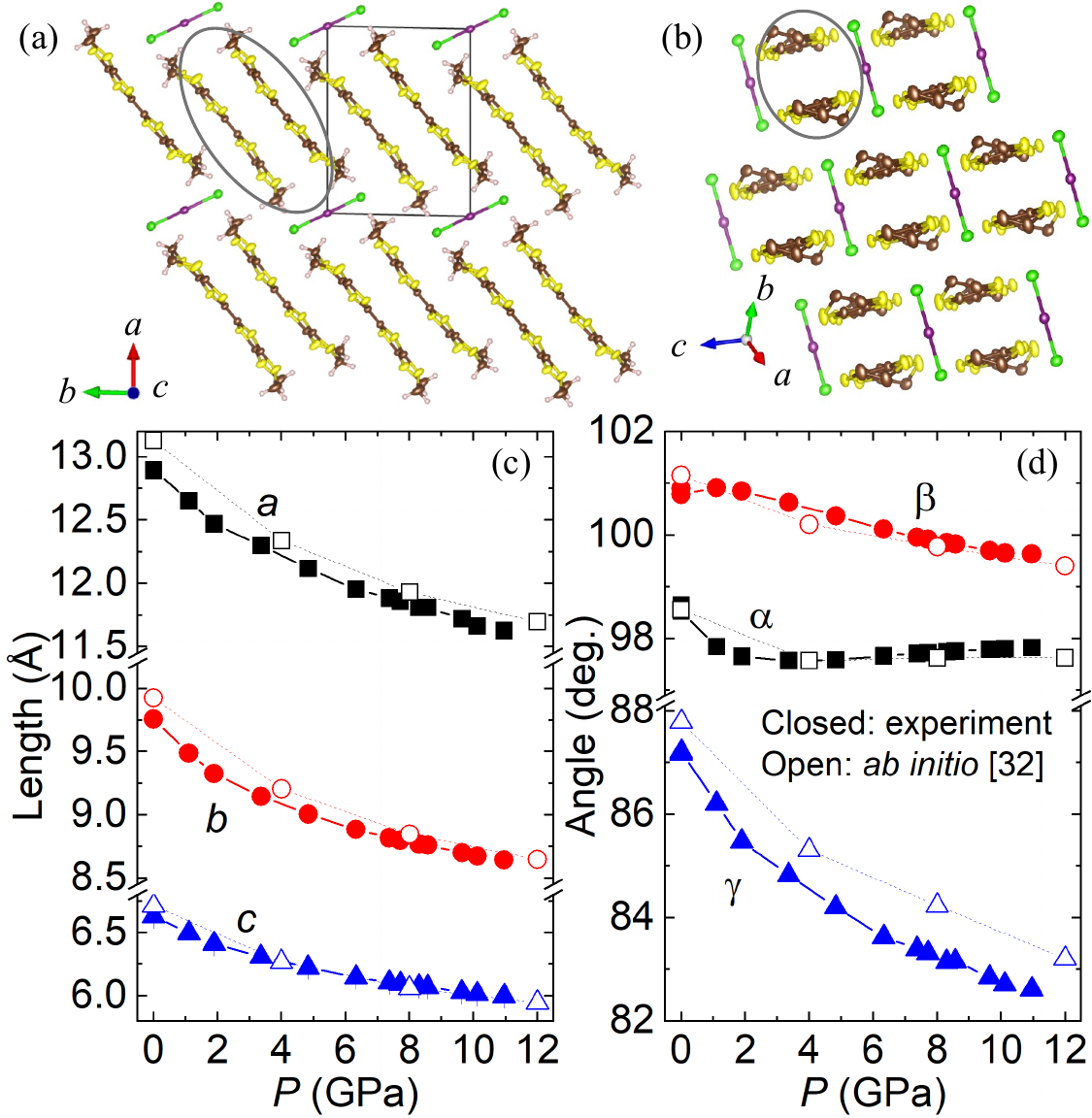}
\caption{
(a)	Crystal structure of \ICl\ viewed along $c$ axis, exhibiting alternating stacking of donor molecules and anions in the $a$ axis. 
(b) In-plane molecular arrangement of \ICl\ viewed along [111] direction, with hydrogens omitted for clarity. 
The parallelogram and ellipse represent the unit cell and the dimer, respectively.
(c), (d) Pressure dependence of lattice parameters. The closed and open symbols are lattice parameters obtained in the present experiment and those estimated from first-principles calculations based on the structural data at ambient pressure \cite{Miyazaki2003}, respectively.} 
\label{fig-lattice} 
\end{center}
\end{figure}

Figure~\ref{fig-lattice}(a) shows the crystal structure of \ICl\ at ambient pressure, with alternating ET and ICl$_2$ layers in the $a$-axis direction. 
In the ET layer, two molecules dimerize and the dimers stack along the $b$-axis direction [Fig.~\ref{fig-lattice}(b)].
With increasing pressure, all lattice lengths shorten monotonically without a phase transition, as shown by the closed symbols in Fig.~\ref{fig-lattice}(c).
When pressure is applied up to $10$~GPa, the $b$-axis length decreases by $12$\% while the $a$- and $c$-axes lengths decrease by approximately $10$\%, indicating anisotropic compression (see the Supplemental Materials \cite{Supplement} and also Refs.~\cite{Birch1947,Mori1998} therein, for pressure-induced variations in the normalized lattice lengths and volume). 
The easy compression along the stacking direction of donor molecules was also observed in other molecular conductors \cite{Watanabe2004,Pashkin2009,Montisci2023}, and the bulk modulus is comparable to that of TMTSF salts \cite{Pashkin2009,Supplement}, suggesting that crystals are compressed in a similar way in organic conductors, regardless of the donor molecule.

In the pressure dependence of the lattice angles [Fig.~\ref{fig-lattice}(d)], a steep decrease in $\alpha$ and a maximum in $\beta$ at $\sim 1$~GPa were observed. By further applying a higher pressure, all angles show a monotonic change. 
Anomalous changes in the lattice angles in the low-pressure region are newly identified in this study. 
This might be related to the changes in the magnetic structure indicated by muon spin rotation and nuclear magnetic resonance measurements \cite{Satoh2009,Eto2010}.

Atomic positions under pressure could be determined with isotropic atomic displacement parameters (see crystal structure data in Supplemental Materials \cite{Supplement}). 
The success of high-pressure single-crystal structural analysis using a laboratory X-ray source will facilitate its applications to organic crystals sensitive to pressure. The obtained data enabled us to discuss the pressure dependence of molecular displacements and electronic structures precisely, and the differences from theoretical studies \cite{Miyazaki2003} are evident as described below.
Note that molecular deformation by applying pressure was hardly observed within the experimental error.
These results indicate that intermolecular compression is more significant than intramolecular compression, and support the validity of the rigid body approximation even up to $10$~GPa that has been used in the high-pressure structural analyses of molecular solids \cite{Filhol1981,Guionneau1996,LePevelen1999,Watanabe2004,Kondo2009}.

{\it Ab initio derivation models---.}~
For the experimental crystal structures, density functional theory (DFT) calculations were performed using \texttt{Quantum Espresso (version 6.7)}\cite{QE}. We utilized norm-conserving pseudopotentials based on the Vanderbilt formalism with plane-wave basis sets as described in \cite{Hamann_ONCV2013, Schlipf_CPC2015} and the generalized gradient approximation by Perdew, Burke, and Ernzerhof  \cite{GGA_PBE} as the exchange-correlation.
We specified cutoff energies of $70$ Ry for plane waves and $280$ Ry for charge densities.
A $3\times 5\times 5$ uniform $\bm{k}$-point mesh with Gaussian smearing was employed throughout the self-consistent iterations.
To derive the model parameters, we used the constrained random phase approximation (cRPA) method~\cite{PhysRevB.70.195104, Imada_JPSJ2010}. Notably, previous studies that applied the cRPA method to molecular solids demonstrated good agreement between the results for the derived effective Hamiltonians and experimental data~\cite{Shinaoka2012, PhysRevResearch.3.043224, Ido2022, PhysRevB.107.L041108, PhysRevLett.131.036401, itoi2024combined}.
We constructed maximally localized Wannier functions (MLWFs) based on the DFT electronic states and determined the parameters using \texttt{RESPACK}~\cite{RESPACK}. In these calculations, the energy cutoff for the dielectric function was set at $3$ Ry.

\begin{figure}[t] 
\begin{center} 
\includegraphics[width=1\columnwidth]{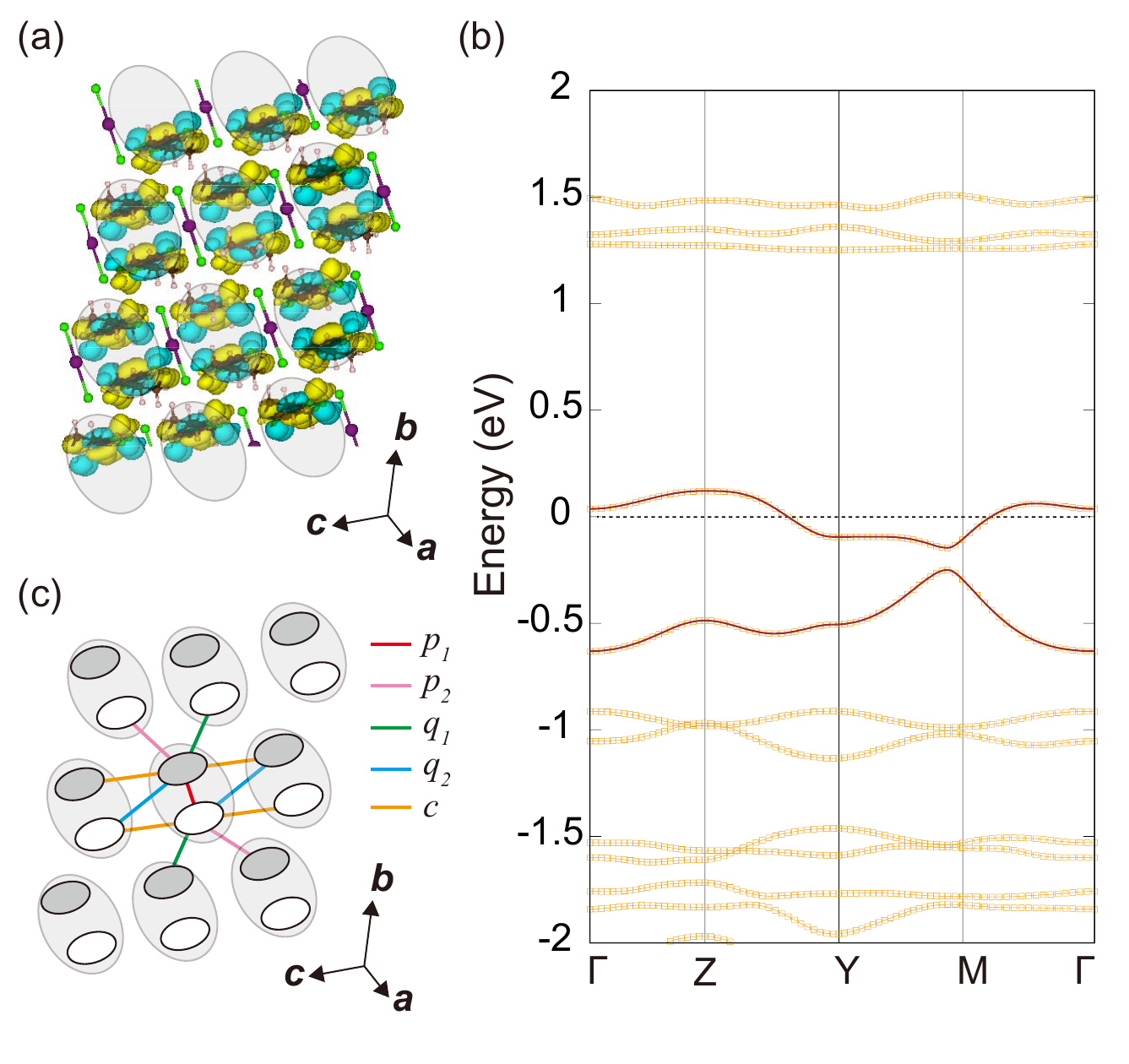}
\caption{
(a) Illustrations of crystal structure with MLWFs of \ICl\ at ambient pressure using \texttt{VESTA}~\cite{VESTA}. The grey ellipse represents a dimer consisting of two MLWFs.
(b) Band structure of \ICl\ at ambient pressure. The band structure obtained by the DFT calculation is denoted by orange symbols, and the tight-binding bands obtained using the MLWFs are denoted by the bold brown lines. 
We set the Fermi energy to zero (the dashed line). 
Here, $\Gamma = (0, 0, 0)$, $Z = (0, 0, \pi)$, $Y = (0, \pi, 0)$, and $M=(0, \pi, \pi)$.
(c) Schematic \textit{ab-initio} model and definition of the transfer integrals. 
} 
\label{fig-bands} 
\end{center}
\end{figure}

Figures \ref{fig-bands}(a) and \ref{fig-bands}(b) show the crystal structure with MLWFs and the band structure of \ICl\ at ambient pressure. 
The two bands near the Fermi energy correspond to the bonding and anti-bonding states of the highest occupied molecular orbitals on the two ET molecules. The energy gap between the upper and lower bands indicates that a dimer state is formed at ambient pressure. The dimers consisting of two MLWFs are represented as the grey ellipses in Fig.~\ref{fig-bands}(a). 
Figure~\ref{fig-bands}(c) shows a schematic of the \textit{ab-initio} model and the definition of the bonds between MLWFs. By employing these MLWFs, we calculated both the transfer integrals and the density--density interactions, namely the on-site and off-site Coulomb interactions. 

\begin{figure*}[t] 
\begin{center} 
\includegraphics[width=2\columnwidth]{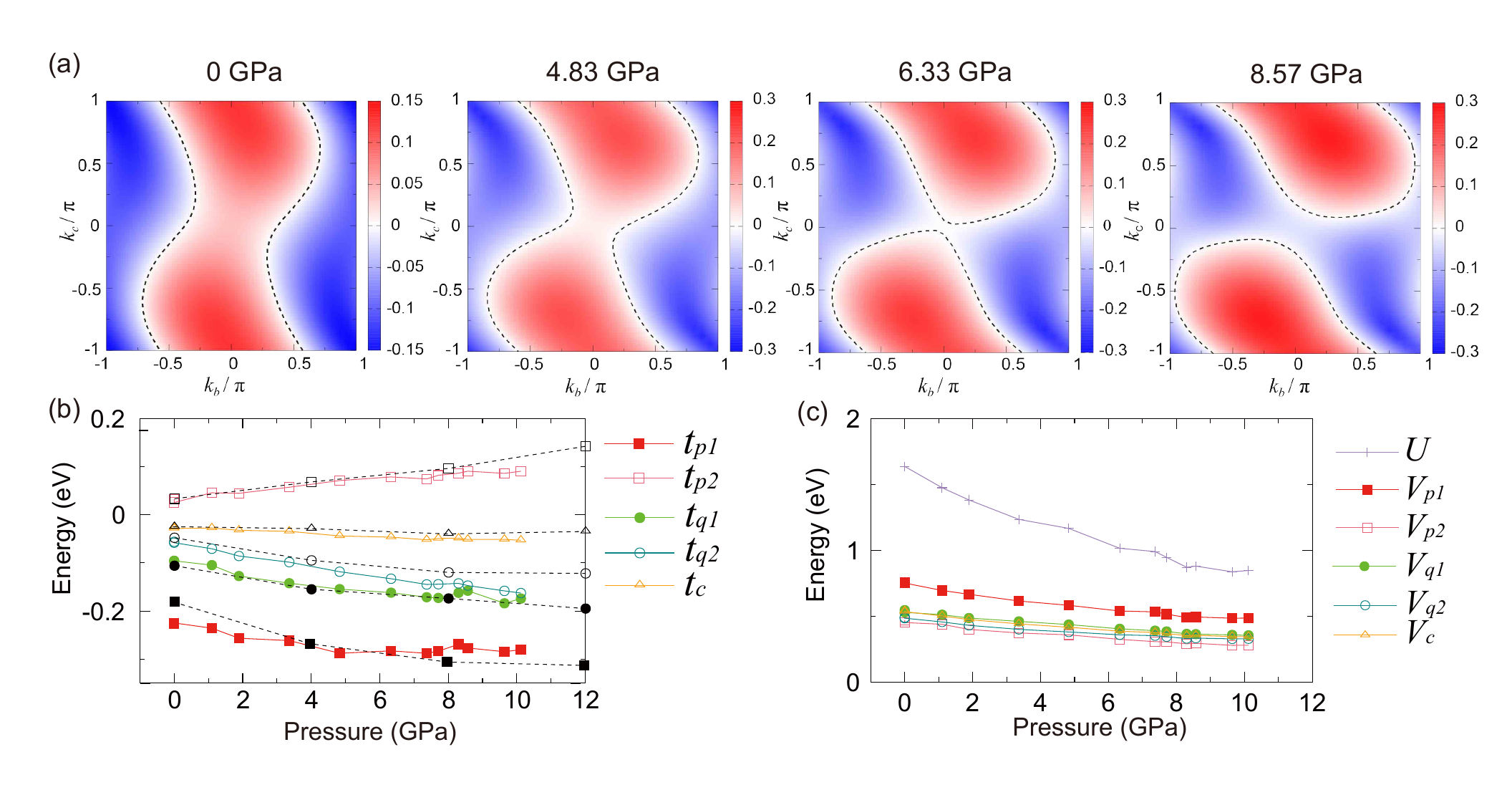}
\caption{Pressure dependence of (a) energy dispersion, with Fermi energy as energy origin, (b) transfer integrals, and (c) Coulomb interactions of \ICl. 
The dashed lines in (a) correspond to the lines where the energy becomes zero, while the black symbols in (b) correspond to transfer integrals obtained in previous studies \cite{Miyazaki2003}.}
\label{fig-params} 
\end{center}
\end{figure*}

Figure~\ref{fig-params}(a) shows the pressure dependence of the energy dispersions with the Fermi energy as the energy origin, for the non-interacting system \cite{fermiplot}. The red and blue shaded areas correspond to positive and negative energies, respectively. 
The dashed lines correspond to the lines of zero energy. The Fermi surface is quasi-one-dimensional at ambient pressure, with two curves in the $k_c$ direction. 
As pressure increases, the two curves gradually approach each other at the $\Gamma$ point, and eventually above $6.33$~GPa, the Fermi surface has a two-dimensional shape.

We also show the pressure dependence of the transfer integrals $t_i$ in Fig.~\ref{fig-params}(b). 
Here, $i$ is the bond index and the correspondence is shown in Fig.~\ref{fig-bands}(c).
The magnitude of the transfer integrals except for $t_{p1}$ tends to increase up to $10$~GPa while $t_{p1}$ seems to be saturated above $6$~GPa, indicating the decrease in the dimerization at higher pressure. This behavior is consistent with the molecular displacement with pressure as discussed in the Supplemental Materials \cite{Supplement}. 
Here, the black symbols indicate the transfer integrals based on the predicted structure under pressure obtained by first-principles calculations in a previous study \cite{Miyazaki2003}.
The difference between our results and the previous study will be discussed later.

The pressure dependence of the screened Coulomb interactions is shown in Fig.~\ref{fig-params}(c). $U$ and $V_i$ represent the screened on-site and off-site Coulomb interactions with the bond index $i$, respectively.
Since the bandwidth broadens when pressure is applied, the screening effect becomes large. Thus, the screened Coulomb interactions decrease monotonically. 
In particular, around $10$ GPa, the magnitude of $U$ halves compared to that at ambient pressure.

{\it Discussion--.}
Combining structural analyses under pressure and \textit{ab initio} calculations, we demonstrate that the Fermi surface drastically changes with a dimensionality change from one to two by applying pressure, i.e. a Lifshitz transition \cite{Lifshitz1960} occurs around $6$ GPa as shown in Fig.~\ref{fig-PT}(a) and Fig.~\ref{fig-params}(a).
To show the difference between the crystal structures obtained by experiments and the first-principles calculations, lattice constants under pressure calculated by Miyazaki \textit{et al.} \cite{Miyazaki2003} are plotted in Figs.~\ref{fig-lattice}(c) and \ref{fig-lattice}(d) with open symbols. 
These predictions are in reasonable agreement with the experimental results, but there is a significant difference in the topology of the Fermi surface.

As seen in Fig.~\ref{fig-params}(b), the overall trend of the pressure dependence of the transfer integrals is almost the same between the experiments and predictions. 
However, there is a difference in that there are slight deviations in the magnitude of the transfer integral, \textit{e.g.} $t_{q2}$ and $t_c$ are slightly larger in the present results.
We showed in Supplemental Materials that these differences are not due to differences in computational methods such as pseudopotentials or ${\bm k}$-point mesh size, but rather to differences between the experimentally determined structure under pressure and the optimized structure by relaxing the internal coordinates under pressure based on the ambient-pressure structure \cite{Supplement}. We also identified differences in molecular displacement that could be responsible for the differences in transfer integrals \cite{Supplement}.

A characteristic feature of this system is that at ambient pressure, the $\Gamma$ point is near the Fermi energy, as seen in Figs.~\ref{fig-bands}(b) and \ref{fig-params}(a). 
As a result, small deviations from the simulated values are expected to significantly impact the electronic state.
The importance of determining atomic positions experimentally is suggested by the fact that in such delicate cases, even if lattice constants and other parameters appear to agree with first-principles calculations, subtle deviations in atomic positions can accumulate and have significant physical consequences. 

In experiments, it has been reported that the charge gap decreases while the N\'{e}el temperature increases with pressure in the AFI phase \cite{Tajima2008,Eto2010} [Fig.~\ref{fig-PT}(b)]. 
This trend seems reasonable since the transfer integrals increase and $U$ decreases with increasing pressure, such that the electron correlation becomes weaker while the magnetic exchange interaction $J_{i}=4t_{i}^2/U$ increases.

The pressure--temperature phase diagram shown in Fig.~\ref{fig-PT}(b) was obtained by electrical resistivity measurements using a cubic anvil system \cite{Taniguchi2003a}. 
Although the pressure application method differs from the present one, hydrostatic conditions were maintained in both experiments, allowing a quantitative comparison.
As can be seen from the phase diagram, the Mott insulating state appears to melt abruptly around $6$~GPa \cite{Taniguchi2003a}. 
This value of pressure is in close agreement with the pressure near the Lifshitz transition, where the Fermi surface shows the dimensional crossover [Fig.~\ref{fig-PT}(a)], suggesting that the Lifshitz transition may affect the melting of the Mott insulator.

In terms of the physical properties in the intermediate temperature range where the crossover occurs, it is expected that the Mott and the Lifshitz transitions are expected to be cooperative. This is because, as the dimensionality increases due to the Lifshitz transition, the charge gap becomes smaller, the bandwidth becomes larger, and the effect of electron correlation is expected to weaken. On the other hand, related fluctuations occur in the vicinity of the Lifshitz transition. How these fluctuations affect systems with strong electron correlations is beyond the scope of this study, but we believe it will be an interesting topic for future work.

A previous study \cite{Nakano2006} showed that the Lifshitz transition occurs around $15.5$~GPa through a linear extrapolation of the pressure dependence of the transfer integrals obtained by Miyazaki \textit{et al}. \cite{Miyazaki2003}. 
When $U$ was around $1.0$~eV, $T_{\rm c}$ became approximately $8$~K at $16$~GPa determined using a two-orbital model with the fluctuation-exchange approximation \cite{Nakano2006}. 
This study identified the importance of improved nesting and an increased density of states near the Lifshitz transition as responsible for the high $T_{\rm c}$. 
However, the discrepancy in the experimentally observed SC critical pressure and the quantitative uncertainty of the $U$ values under pressure were problematic.  
In our analysis, the Lifshitz transition appears around $6$~GPa, above which $U$ saturates around $1.0$~eV.
Considering the similarities of the Fermi surface geometry and $U$ values to the situation presented in the previous studies, a similar mechanism could reproduce the SC state under pressure in quantitative agreement with the experiment.
The ability to derive electron correlation effects for organic materials, which are known to exhibit significant pressure dependence \cite{itoi2024combined,Kato2024}, without contradicting previous studies, is noteworthy. 
The fact that these effects could be quantitatively derived from first principles represents a major advancement in the field of organic materials research and is expected to contribute to future material design and the discovery of new materials.

In systems such as TMTSF salts, it has been reported that \cite{Jerome1980,Ducasse1986,LePevelen2001} applying pressure to a quasi-one-dimensional system increases the dimensionality and weakens the electron correlation, leading to the disappearance of charge- and spin-ordered phases and the emergence of superconductivity due to fluctuations near the transition point. 
However, there are not many reports of a complete transition to a two-dimensional system. 
Our study has shown that in quasi-one-dimensional systems, the emergence of superconductivity is linked to pressure-induced dimensional changes and a marked decrease in electron correlation. 
In such systems, there is potential for the emergence of new phenomena involving fluctuations related to the Lifshitz transition and electron correlation effects. 
In particular, the pressure effect is pronounced in organic conductors, suggesting many candidate materials for similar transitions.

{\it Summary--.}
This paper investigates the changes in the Fermi surfaces of \ICl\ under pressure by combining structural analysis and first-principles calculations. 
We demonstrate that the Fermi surface changes from a one-dimensional to a two-dimensional structure, i.e. a Lifshitz transition exists at around $6$ GPa, at which the Mott insulator--metal transition and the SC transition have been experimentally observed.
The pressure application significantly reduces the electronic correlations by enhancing the bandwidth and suppressing $U$, which can explain the decrease in the charge gap and the increase in the N\'{e}el temperature with increasing pressure. 
A comparison with a previous study \cite{Nakano2006} indicates that the occurrence of the SC state is consistent with the experimentally obtained phase diagram.
In addition, this study reveals that \textit{ab initio} analyses based on experimentally determined structures are indispensable for achieving high-precision analyses of organic conductors, including SC transitions. 
We believe that these results expand the possibilities for material design and will open the door to the discovery of new phenomena in organic conductors.

\begin{acknowledgements}
We wish to thank T. Hattori, T. Kato, T. Tsumuraya, and T. Misawa for their helpful contributions. 
We also appreciate N. Hamaya and A. Ohmura for their involvement in our early work on structural analysis under pressure.
This work was supported by the Japan Society for the Promotion of Science KAKENHI Grants No. 21H01041, No. 21K03438, No. 22K03526, and 23H01788.
Crystallographic information files for the \ICl\ at each pressure are available in the Supplemental Materials \cite{Supplement}.
The computation in this work was performed using the facilities of the Supercomputer Center, Institute for Solid State Physics, University of Tokyo.
T.~K. and K.~Y. contributed to this work equally.
\end{acknowledgements}

%\bibliography{aps}
%

\end{document}

\title{Supplemental Materials for ``Pressure-Induced Topological Changes in Fermi Surface of Two-Dimensional Molecular Conductor"}
\author{T.~Kobayashi}
 \email{tkobayashi@mail.saitama-u.ac.jp}
 \affiliation{Graduate School of Science and Engineering, Saitama University, Saitama 338-8570, Japan}
 \affiliation{Research and Development Bureau, Saitama University, Saitama 338-8570, Japan}
  \author{K.~Yoshimi}
 \email{k-yoshimi@issp.u-tokyo.ac.jp}
  \author{H.~Ma}
 \affiliation{Institute for Solid State Physics, University of Tokyo, Kashiwa, Chiba 277-8581, Japan}
 \affiliation{Beijing National Laboratory for Condensed Matter Physics and Institute of Physics, Chinese Academy of Sciences, Beijing 100190, China}
 \author{S.~Sekine}
  \affiliation{Graduate School of Science and Engineering, Saitama University, Saitama 338-8570, Japan}
\author{H.~Taniguchi}
 \affiliation{Graduate School of Science and Engineering, Saitama University, Saitama 338-8570, Japan}
\author{N.~Matsunaga}
\author{A.~Kawamoto}
 \affiliation{Department of Condensed Matter Physics, Graduate School of Science, Hokkaido University, Sapporo 060-0810, Japan}
  \author{Y.~Uwatoko}
 \affiliation{Institute for Solid State Physics, University of Tokyo, Kashiwa, Chiba 277-8581, Japan}
\date{\today}% It is always \today, today,
\maketitle

\beginsupplement

\section{1.~Pressure dependence of normalized lattice parameters}
\label{secA}
\begin{figure}[t] 
\begin{center} 
\includegraphics[width=0.8\columnwidth]{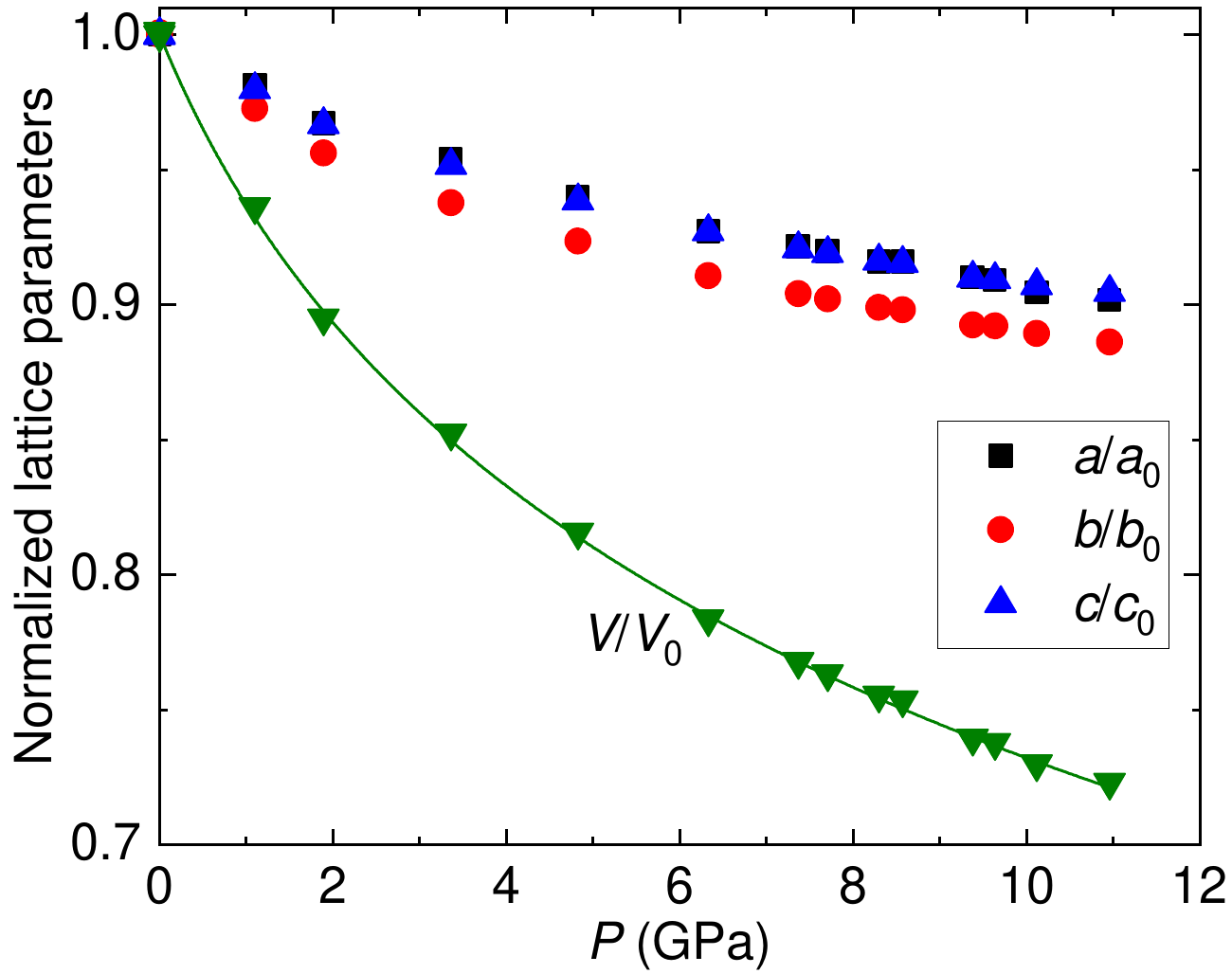}
\caption{
(a) Pressure dependence of normalized lattice lengths and volume of \ICl. 
The solid line is a fitting curve using Eq.~(\ref{BM}).
} 
\label{fig-normlattice} 
\end{center}
\end{figure}
The pressure dependence of lattice parameters and volume normalized by their respective values at ambient pressure is shown in Fig.~\ref{fig-normlattice}. $a/a_0$ and $c/c_0$ show almost the same pressure dependence, whereas $b/b_0$ is more compressed.

The relationship between volume and pressure is known to be described by the Birch-Murnaghan equation of state \cite{Birch1947},
\begin{equation}
    P(V) = \frac{3}{2}B_0 \left[ \left( \frac{V_0}{V} \right)^{\frac{7}{3}} - \left( \frac{V_0}{V} \right)^{\frac{5}{3}} \right] \left\{ 1 + \frac{3}{4} \left( B'_0 - 4 \right) \left[ \left( \frac{V_0}{V} \right)^{\frac{2}{3}} - 1 \right] \right\}
    \label{BM}
\end{equation}
%\begin{multline}
%    P(V) = \frac{3}{2}B_0 \left[ \left( \frac{V_0}{V} \right)^{\frac{7}{3}} - \left( \frac{V_0}{V} \right)^{\frac{5}{3}} \right]\\
%    \times \left\{ 1 + \frac{3}{4} \left( B'_0 - 4 \right) \left[ \left( \frac{V_0}{V} \right)^{\frac{2}{3}} - 1 \right] \right\}
%    \label{BM}
%\end{multline}
where $B_0$, $B_{0}'$, and $V_0$ are the bulk modulus, its first derivative with respect to pressure, and the unit-cell volume at ambient pressure, respectively. 
The solid line in Fig.~\ref{fig-normlattice} is a fit using Eq.~(\ref{BM}), and the relationship between volume and pressure is well reproduced by $B_0 = 12.60 \pm 0.88$~GPa and $B_{0}' = 6.08 \pm 0.38$. 
These values are comparable to those for other organic conductors \cite{Pashkin2009}.

\section{2.~Changes in molecular positions with pressure}
\label{secB}
In \ICl, a unit cell contains two ET molecules that are connected through inversion symmetry, resulting in two distinct overlapping patterns in the stacking direction of the ET molecules: intradimer and interdimer. Furthermore, the molecules stack parallel to each other, which is maintained even under pressure. The relationship between the overlap integrals and molecular positions in such a case has been systematically investigated by Mori \cite{Mori1998}. 
We here discuss the pressure variation of the molecular position based on this study. 

\begin{figure*}[t] 
\begin{center} 
\includegraphics[width=\columnwidth]{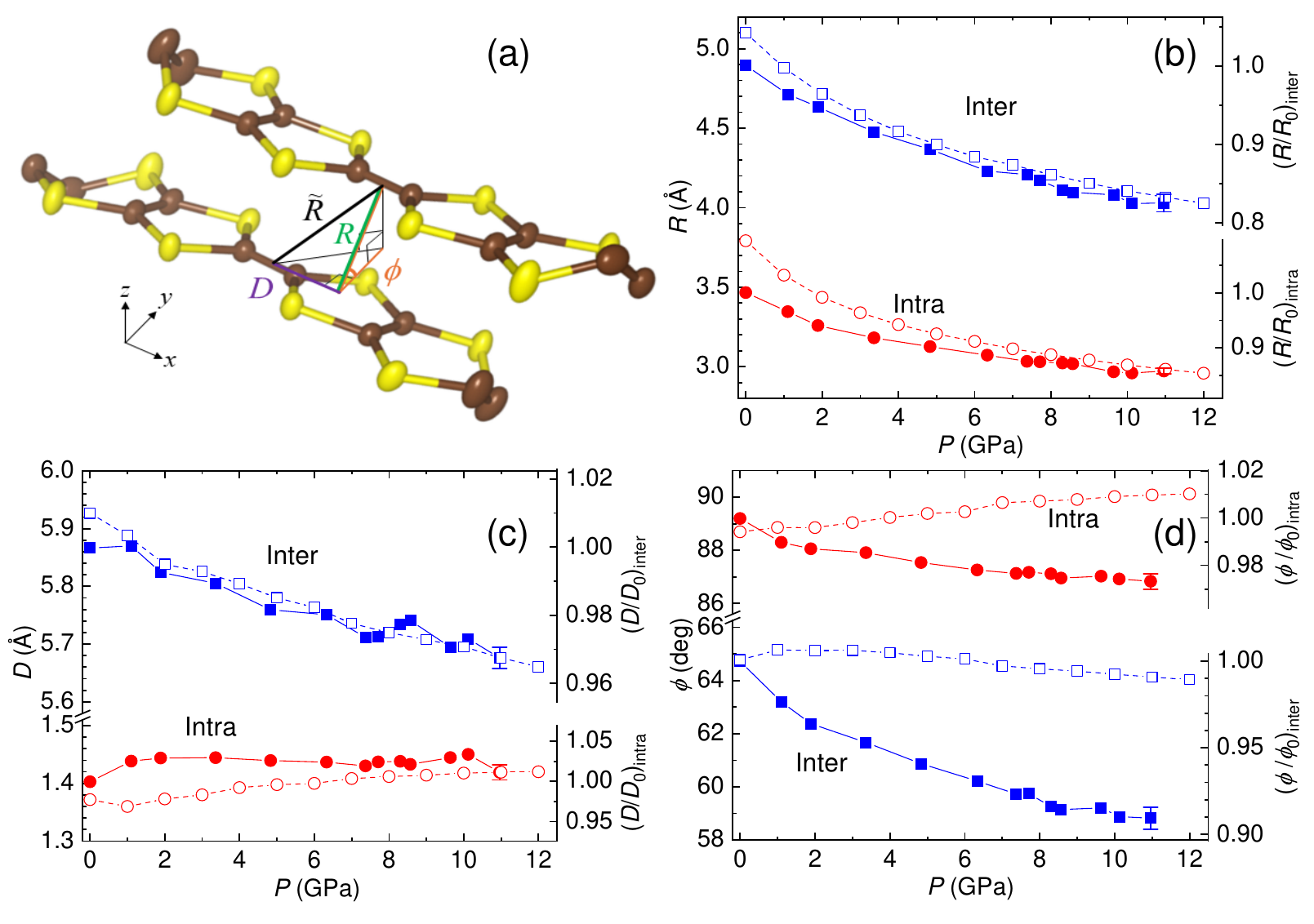}
\caption{
(a) Definitions of $R$, $D$, and $\phi$ for discussing the relative distance between two ET molecules. (b)-(d) Pressure variations of $R$, $D$, and $\phi$ between ET molecules along the $b$ axis in \ICl, which is the stacking direction of ET molecules.
Red circles and blue squares indicate intradimer and interdimer relationships, respectively. 
Closed and open symbols are results obtained from experiments and first-principles calculations, respectively. 
The error bars for the experimental data at the measured maximum pressure are comparable to those under other pressures. 
The right axis shows the ratio of each parameter with respect to the value at ambient pressure. 
} 
\label{fig-mol_distance} 
\end{center}
\end{figure*}

We define the average position of an ET molecule as the centroid of the ten atoms forming the planar TTF unit (C$_3$S$_2$)$_2$. The vector connecting these centroids between two ET molecules is denoted as $\tilde{\bm R}$, with its magnitude represented by $\tilde{R}$. Here, we introduce the following coordinate system: the molecular long axis as the $x$ axis, the molecular short axis as the $y$ axis, and the direction perpendicular to the molecular plane as the $z$ axis [Fig.~\ref{fig-mol_distance}(a)]. The length of the projection of $\tilde{R}$ along the $x$ axis is denoted as $D$, which allows us to examine the displacement between two molecules along the molecular long axis. We then consider the root mean square of the projections of $\tilde{R}$ along the $y$ and $z$ axes, represented as $R$. This provides a measure of the variation in distance between molecules. Additionally, the angle $\phi$ formed between the molecular plane and $R$ (where $\tan \phi = z/y$) is also a significant parameter, closely correlated with the overlap integral. 

Figures~\ref{fig-mol_distance}(b)--(d) display the pressure dependence of $R$, $D$, and $\phi$, respectively. At ambient pressure, $R$ and $D$ within dimers (red closed circles) are significantly smaller than those between dimers (blue closed squares). According to Mori's definition \cite{Mori1998}, the intradimer overlap ($\phi = 89^{\circ}$) is in the ring-over-bond (RB) mode, while the interdimer overlap ($\phi = 65^{\circ}$) is in the ring-over-atom (RA) mode. 
With applied pressure, inter- and intra-$R$  decrease by $18$\% and $14$\%, respectively. 
The displacement along the long axis $D$ exhibits qualitatively different pressure-induced changes within and between dimers. 
With increasing pressure, intra-$D$ remains relatively unchanged and inter-$D$ decreases although their changes are approximately $3$\%. 
Referring to Fig.~2(b) in Ref.~\cite{Mori1998}, the overlap integral increases when $D$ changes from $5.9$~\AA\ to $5.7$~\AA\ in the RA overlap mode ($\phi = 60^{\circ}$). 
These results indicate that the increasing ratio of interdimer overlap integrals becomes greater than that of intradimer ones.

Regarding the angle $\phi$ shown in Fig.~\ref{fig-mol_distance}(d), we observed more significant changes between dimers compared to within them. Referring to Fig.~2(a) in Ref.~\cite{Mori1998}, the RA overlap mode shows a maximum at $\phi \sim 59^{\circ}$, while the RB mode exhibits a minimum (maximum in magnitude) at $\phi = 90^{\circ}$. In our results, an increase in pressure leads to a decrease in intra-$\phi$ from $89^{\circ}$ to $87^{\circ}$, which tends to reduce the overlap integral. Conversely, a change in inter-$\phi$ from $65^{\circ}$ to $59^{\circ}$ suggests that the overlap integral is approaching its maximum. Thus, the pressure-induced variations in $R$, $D$ and $\phi$ reveal that the overlap integral increases more substantially between dimers than within them, suggesting that pressure application reduces the dimerization. 

\section{3.~Analysis of First-Principles Calculations: Pseudopotentials, k-point Mesh, and Structural Relaxation}
\label{secC}
In this section, we investigate the effect of different pseudopotentials, $\bm{k}$-point meshes, and structural relaxations on the results of first-principles calculations, mainly focusing on the band structure and Lifshitz transition. 
% The following sections provide a detailed analysis of each of these effects. 
All calculations were performed by using \texttt{Quantum Espresso (version 6.7)} \cite{QE} using the generalized gradient approximation (GGA) by Perdew, Burke, and Ernzerhof (PBE) \cite{GGA_PBE} as the exchange-correlation function, with plane wave and charge cut-off energies set to $70$ and $280$ Ry, respectively. Gaussian smearing was employed throughout the self-consistent iterations.

\begin{figure*}[t] 
\begin{center} 
\includegraphics[width=\columnwidth]{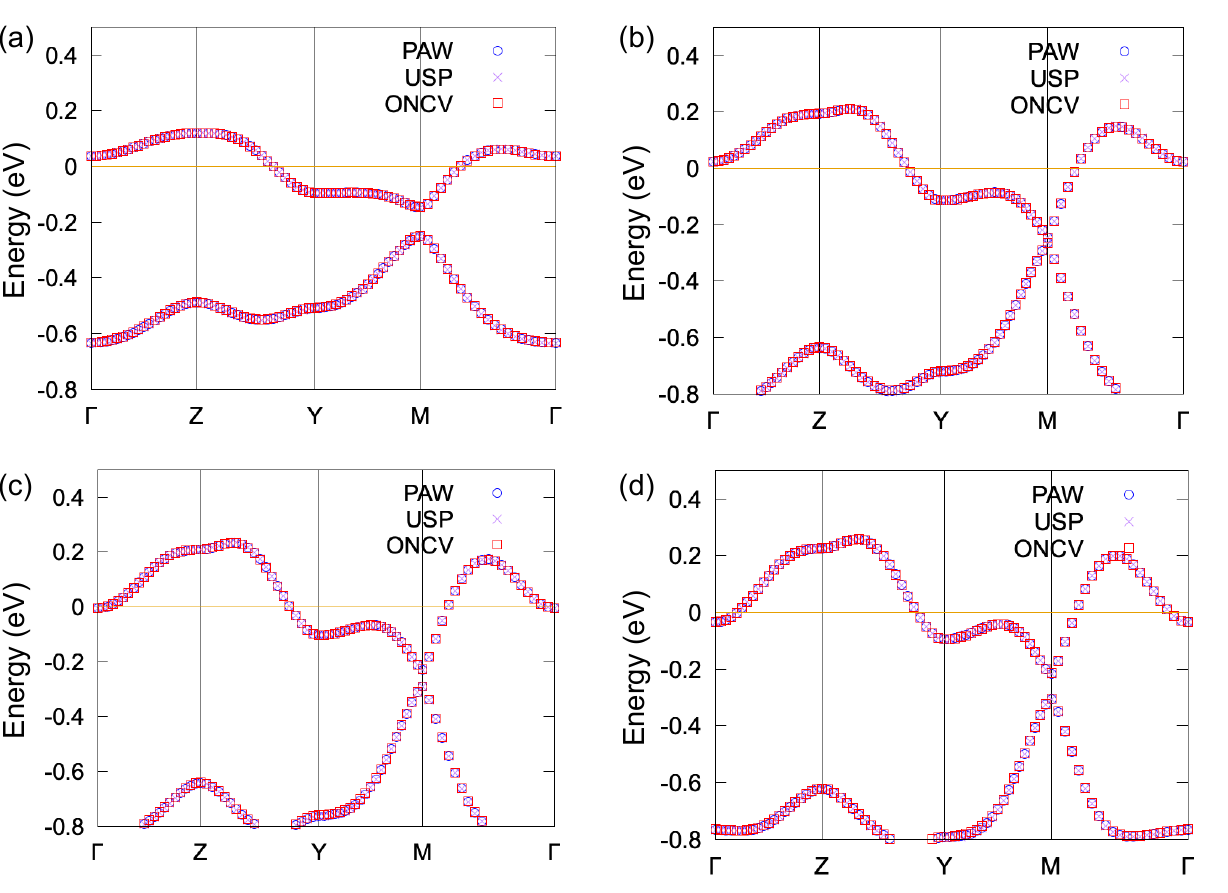}
\caption{
Band structures obtained by USP,  PAW, and ONCV at (a) 0 GPa, (b) 4.83 GPa, (c) 6.33 GPa, and (d) 8.30 GPa using the experimentally obtained crystal structures.
} 
\label{fig-band_pseudo} 
\end{center}
\end{figure*}

First, band structure calculations were performed on a $3 \times 5 \times 5$ uniform $\bm{k}$-point mesh for three different types of pseudopotentials (Ultra Soft Potential (USP), Projector Augmented Wave (PAW), and Optimised Norm-Conserving Vanderbilt (ONCV)) with the experimentally determined crystal structure under pressure. 
Figure~\ref{fig-band_pseudo} shows the band structures obtained at four different pressures: $0$~GPa, $4.83$~GPa, $6.33$~GPa, and $8.30$~GPa. 
Despite slight variations in numerical precision, the overall band dispersion and Fermi surface topology were consistent across all three pseudopotentials. 
This suggests that the choice of pseudopotential does not significantly influence the qualitative features of the band structure, particularly the energy levels near the Fermi surface, and thus we use the ONCV potential in the following analyses. 

\begin{figure*}[t] 
\begin{center} 
\includegraphics[width=\columnwidth]{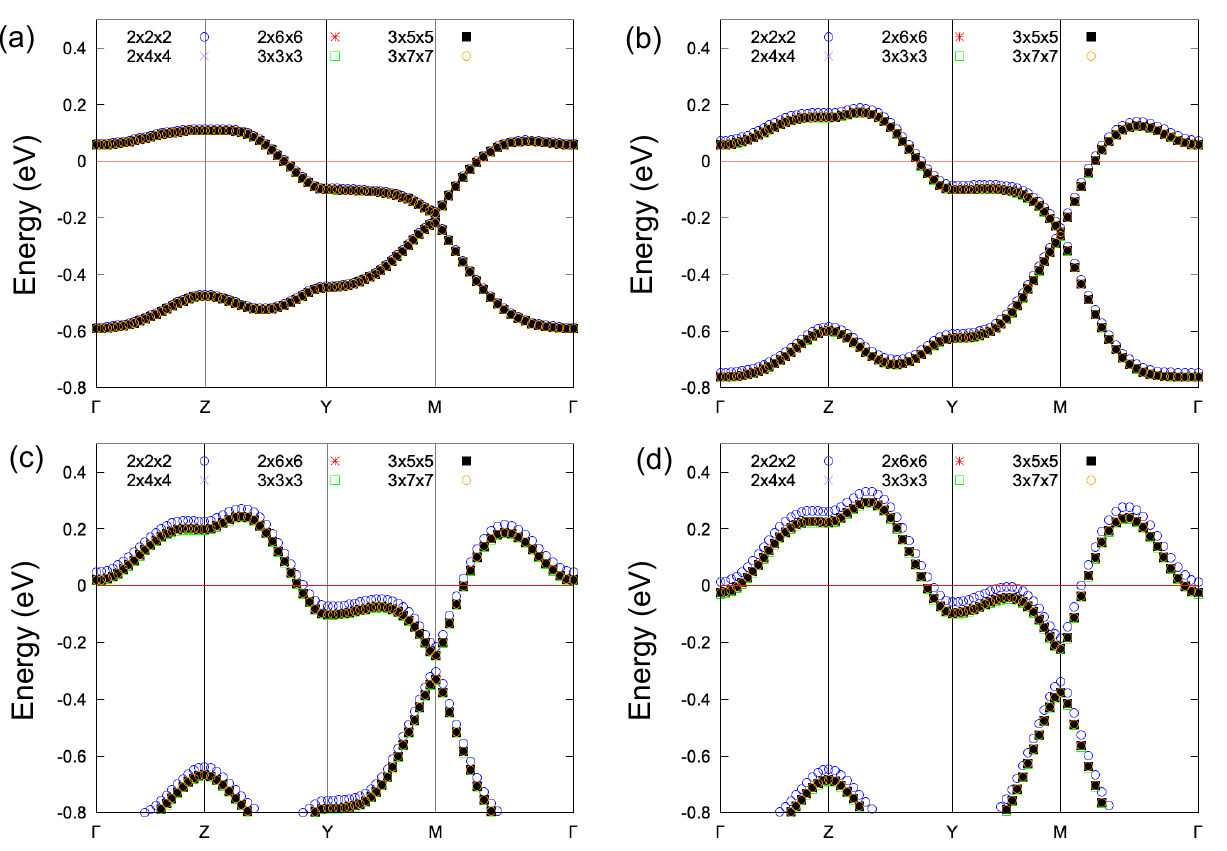}
\caption{
Band structures obtained using uniform $\bm{k}$ meshes of $2 \times 2 \times 2$, $2 \times 4 \times 4$, $2 \times 6 \times 6$, $3 \times 3 \times 3$, $3 \times 5 \times 5$, and $3 \times 7 \times 7$ at (a) $0$~GPa, (b) $4$~GPa, (c) $8$~GPa, and (d) $12$~GPa. The calculations were carried out using the crystal structures obtained from the structural relaxation calculation fixed to the lattice structures reported in the previous study \cite{Miyazaki2003}.
} 
\label{fig-band_mesh} 
\end{center}
\end{figure*}

Next, we investigate the convergence of $\bm{k}$-point mesh size on the band structure; we performed band structure calculations by varying $\bm{k}$-point meshes.
First, we performed structural relaxation calculations using a $3\times5\times5$ ${\bm k}$-point mesh for each of the following pressures: $0$~GPa, $4$~GPa, $8$~GPa, and $12$~GPa.
Using the obtained structure, we set the ${\bm k}$ mesh to $2\times2\times2$, $2\times4\times4$, $2\times6\times6$, $3\times3\times3$, $3\times5\times5$, and $3\times7\times7$, and calculated the band structure for each mesh.
As shown in Fig.~\ref{fig-band_mesh}, we confirmed that the results for the coarse $2\times2\times2$ mesh were different from those for other meshes, but that the results for finer meshes (larger than $2\times4\times4$) were almost the same.
The Lifshitz transition is characterized by the appearance of a band crossing the Fermi energy near the $\Gamma$-point, which alters the topology of the Fermi surface. 
Our results show that a ${\bm k}$-point mesh finer than $2\times4 \times4$ makes little difference in capturing this transition.

\begin{figure*}[t] 
\begin{center} 
\includegraphics[width=\columnwidth]{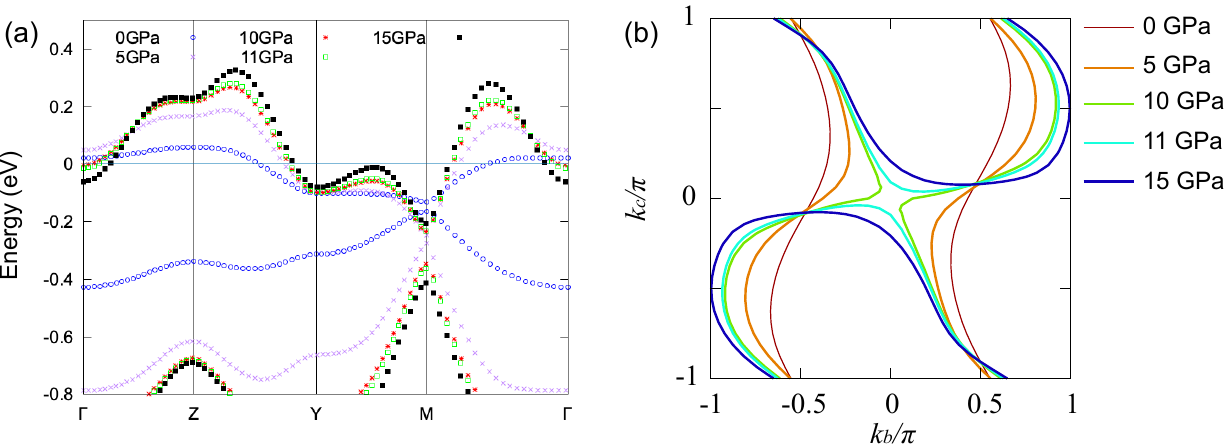}
\caption{
(a) The band structures and (b) the Fermi surface at $0$~GPa, $5$~GPa, $10$~GPa, $11$~GPa, and $15$~GPa using the crystal structures obtained by the structure relaxation calculations.
} 
\label{fig-fermi-pdep} 
\end{center}
\end{figure*}

Finally, we investigated how the crystal structure, band structure, and Fermi surface change when lattice and total atomic structural relaxation calculations are performed under pressure using first-principles calculations.
Starting from the experimentally obtained crystal structure at ambient pressure, we performed structural relaxation calculations in increments of $1$ GPa from $0$ GPa to $15$ GPa.
As the pressure increases, the band energy at the $\Gamma$ point gradually decreases and becomes lower than the Fermi energy, i.e. the Lifshitz transition occurs in the same way as in calculations using experimentally obtained crystal structures.
However, the transition pressure ($\sim 10$ GPa) is higher than the value calculated using the experimental structure (see Fig.~\ref{fig-fermi-pdep}).

\begin{figure}[t] 
\begin{center} 
\includegraphics[width=1\columnwidth]{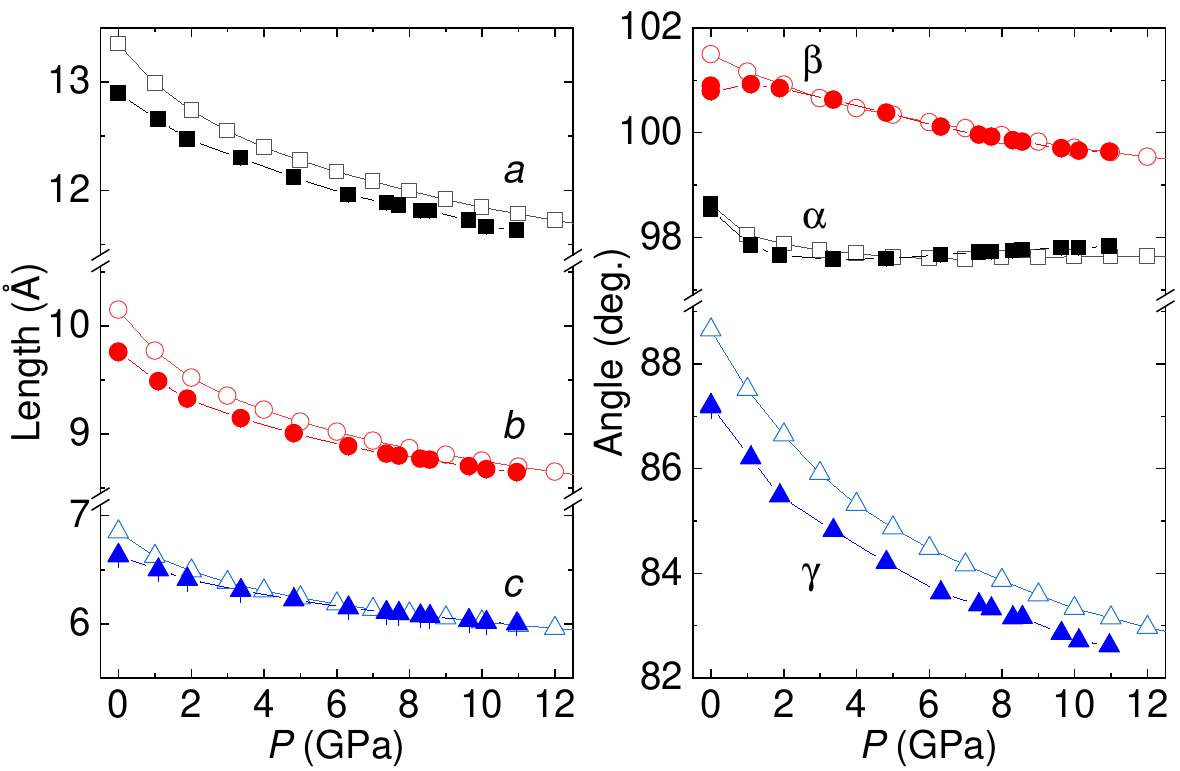}
\caption{
Pressure dependence of (a) lattice lengths and (b) angles obtained from experiments (closed symbols) and first-principles calculations (open symbols).
} 
\label{fig-lattice-pdep} 
\end{center}
\end{figure}

To compare the crystal structures obtained by first-principles calculations with experimental ones, we analyzed the pressure dependence of the lattice information and the arrangement relationship between the ET molecules.
The calculated lattice constants tend to agree with the experimental results, as shown in Fig.~\ref{fig-lattice-pdep}. 
However, the lattice angle $\gamma$ was consistently off, affecting the relative relationship between ET molecules.
To investigate the details, the positions and orientations of the ET molecules obtained from structural relaxation at different pressures were compared with those obtained experimentally.
As shown in Figs.~\ref{fig-mol_distance}(a) and (b), 
the pressure dependence of $R$ and $D$ between the theoretical (open symbols) and experimental (closed symbols) values is qualitatively the same.
However, as seen from Fig.~\ref{fig-mol_distance}(c), a significant deviation in the relative orientation of the ET molecules $\phi$ was found. 
In the experimental structure, the inter- and intra-$\phi$ values decrease with increasing pressure, whereas the crystal structures obtained from the structural relaxation calculation show almost the same values as those at ambient pressure. 
In particular, looking at the experimental results, the inter-$\phi$ decreases more than the intra-$\phi$. 
The shift of the inter-$\phi$ across the unit cell affects the two-dimensionality. It is therefore thought to be the main cause of the difference in the pressure dependence of $t_c$ and $t_{q2}$ between the first-principles calculations using the experimental structures and the structures obtained from the structural relaxation calculations.
The reasons why changes in molecular displacements with pressure cannot be predicted by first-principles calculations are currently unclear. From the viewpoint of theory-driven material design, it is necessary to examine them including van der Waals interactions.

%\begin{figure*}[t] 
%\begin{center} 
%\includegraphics[width=2\columnwidth]{inter_mol_distance_compare.pdf}
%\caption{
%Pressure dependence of (a) $R$, (b) $D$, and (c) $\phi$ between ET molecules obtained from experiments (closed symbols) and first-principles calculations (open symbols).
%} 
%\label{fig-inter_mol_distance_compare} 
%\end{center}
%\end{figure*}

%\bibliography{aps}
%